\documentclass{article}

\usepackage{graphicx}				
\usepackage{amsmath}				
\usepackage{amsfonts}
\usepackage{url}				
\usepackage[colorlinks=true,linkcolor=blue,anchorcolor=black,citecolor=blue,filecolor=black,menucolor=black,urlcolor=blue,breaklinks=true,pdfhighlight=/P,pdfmenubar=true,pdftoolbar=true,pdfpagelabels=true,pdfstartpage=1,pdfstartview=FitV,pdftitle={The Effect of Integrating Travel Time},pdfsubject={Pedestrian, Crowd, Mass, Simulation, Dynamics, VISWALK, VISSIM, FAST},pdfauthor={Kretz},pdfcreator={tobias Kretz},pdfproducer={t kretz},pdfkeywords={Pedestrian, Crowd, Mass, Simulation, Dynamics, VISWALK, FAST}]{hyperref}	
\usepackage[numbers,sort&compress]{natbib}	
\usepackage{hypernat}				
\usepackage[american]{babel}		
\usepackage{simplemargins}
\setallmargins{1.0in}

\begin{document}
%
\title{The Effect of Integrating Travel Time}
\author{Tobias Kretz \\ PTV Group \\ Haid-und-Neu-Stra{\ss}e, D-76131 Karlsruhe\\ \tt{Tobias.Kretz@ptv.de}}

\maketitle

\begin{abstract}
This contribution demonstrates the potential gain for the quality of results in a simulation of pedestrians when estimated remaining travel time is considered as a determining factor for the movement of simulated pedestrians. This is done twice: once for a force-based model and once for a cellular automata-based model. The results show that for the (degree of realism of) simulation results it is more relevant if estimated remaining travel time is considered or not than which modeling technique is chosen -- here force-based vs. cellular automata -- which normally is considered to be the most basic choice of modeling approach.
\end{abstract}

\section{Introduction}
Users of crowd simulation software tools have noted that these tools mostly produce a too high pedestrian density close to a corner when a large group of pedestrians has to walk around this corner compared to real people in the same situation \cite{Rogsch2010}. This leads to considerably increased travel times of simulated pedestrians compared to real people as well as compared to travel times of simulated pedestrians on a straight path of the same length. The same authors express their conviction that real people on contrast organize smarter in such a situation. Real people do not insist to walk a short path, but rather accept a longer walking path in exchange for a moderate local density and therefore can maintain their walking speed and in final consequence achieve a smaller travel time, respectively a travel time which is only slightly increased compared to them walking a straight path with the same length. Concerning a first, qualitative empirical evidence, someone reading this article in an office in a busy CBD area can witness this behavior by looking out the window. For everyone with a less busy window view, YouTube can help \cite{Youtube2012}. 

Even if one had no empirical indications of human walking behavior in such a situation, it would be a good guess that people desire mostly to minimize their travel time and not their travel distance. After all traffic planning for road networks rests on the assumption that one of the most important factors -- if not {\em the} most important one -- for route choice of vehicle drivers is travel time and not route length. The same people who steer vehicles make up for a large share of pedestrians in most situations where pedestrians occur. They may move in different circumstances with different implications of their route and speed choices for their personal effort, but they are the same people.

\section{Integrating Travel Time in(to) Models of Pedestrian Dynamics}
As real people, when we are in a hurry and desire most to minimize time to arrival, we are faced with the problem that we do not know which direction will be the one that allows earliest arrival -- imagine for example to walk through a busy train station while being late in time for the train. Even in retrospective when having arrived it is not always possible to tell if the arrival actually was the earliest possible or if along the path one of the many decisions for a movement direction was wrong with regard to minimizing travel time. This is not only a consequence of the complexity of the problem, but also simply of incomplete information.

Simulating pedestrians we are in a similar situation. If with given constraints for every point in time, every position, and every destination the exact remaining travel time was known for every type of pedestrian, it would in fact not be necessary to simulate anymore. For what would be the additional information one could gain from such a simulation? The simulation would be more or less a discrete representation or instance of the multi-dimensional field of remaining travel times. Furthermore such a simulation would show an optimal solution which most probably would not be a realistic one.

Therefore what is possible is to {\em estimate} a map of remaining travel times in each time step and make pedestrians choose their desired or preferred walking direction according to this map respectively field. The estimation process at first should model realistically the travel time estimation real people do about the direction options they have and second it should be self-consistent with the actual model of pedestrian dynamics in which the travel time estimation module is used.

This is not a simple task and as a consequence models of pedestrian dynamics usually set the direction of the shortest path as the main or preferred walking direction \cite{Schadschneider2009,Schadschneider2009b}. However, in the last few years at least two such combined systems of a remaining travel time estimation and its usage in a pedestrian simulation have been introduced. Both have in common that the travel time estimation (or delay time estimation) is first done for small areas and then numerically integrated to receive the field of remaining travel time from a spot to the destination, which then is used in the pedestrian model where it amends or replaces (the impact of) the field of remaining distance to destination (aka ``static potential'').

The first model is the F.A.S.T. model \cite{Kretz2006f,Kretz2007a,Kretz2006d,Kretz2007b}. The F.A.S.T. model has developed out of a cellular automata approach. It is arguable if it still is a cellular automata \cite{Kretz2010f}, however, the important property that the pedestrians move like chess figures on a regular grid is preserved, making the model both: very fast for execution \cite{Kretz2010b} and rather coarse-grained in the results. A second property is that it is attractiveness of spots which are used in the computation for the next movement step. This distinguishes it from the Social Force Model, where the property which is directly calculated is acceleration (and locations follow via integration over time). The method to estimate travel times considers only the spatial distribution of pedestrians and not their dynamic movement properties\footnote{Of course obstacles are also considered. Considering only them would be no big deal, as they are static. The field of remaining travel times would then be identical or at least very similar to the field of remaining distances. And it would only be necessary to compute it once in advance of the simulation. One of the difficulties of considering travel time is that the field has to be recalculated in every time step or at least frequently.} \cite{Kretz2009}. The method for numerical integration is the so called {\em variant 2} as introduced in \cite{Kretz2010a}. Compared to the Euclidean metric -- if a field of constant local distances or travel times is integrated -- it includes some artifacts, but the way it is used in the pedestrian model reduces them largely and the computation time of {\em variant 2} is below the exact methods mentioned in the next paragraph. In this way a main intention of the F.A.S.T. model -- having a very fast pedestrian simulation -- is preserved, even when the travel time estimation module is active.

The second model is the Social Force Model \cite{Johansson2007,Helbing2009} respectively its implementation in PTV Viswalk \cite{VISSIM2011,Kretz2008b}. In this approach of a combination of a pedestrian dynamics model with a travel time estimation the velocity of a pedestrian is considered when it is estimated which delay time he will impose to another pedestrian in his surrounding \cite{Kretz2011e}\footnote{Also see that paper for an extensive discussion of related work.}. For the numerical integration the Fast Iterative or Fast Marching Method is used. Either method -- in case of distance computation -- yields the nearly correct Euclidean distances \cite{Kimmel1998,Jeong2007,Jeong2007b,Jeong2008}. In the Social Force Model the negative gradients of the resulting field of estimated remaining travel time are used as direction of the desired velocity of a pedestrian being located at the field's corresponding position. Apart from that PTV Viswalk's Social Force Model remains unchanged.

Both models have been used with their travel time estimation module in different use cases \cite{Kretz2009c,Kretz2010c,Kretz2010x,Kretz2011f,Kretz2012c,Youtube2012b}. However, in this contribution it is not intended to solely demonstrate the effect of the usage of the field of remaining travel times, but to argue that the decision to use it or not is more important (i.e. has a greater effect) than the decision which basic modelling approach -- force-based or cellular automata-based -- is applied.

\subsection{Implications and Interpretation}
Normally this would be the place to define the math of the travel time estimation module(s). However, this has been done extensively before \cite{Kretz2009,Kretz2011e}. Instead of a repetition this subsection discusses some implications of the method, i.e. it takes some interpretative steps.

It has been observed \cite{Moussaid2009,Chraibi2010} and sometimes criticized \cite{Steffen2008} that the Social Force Model (and most if not all of its variants) strictly superpose the repulsive forces of pedestrians. Superposition means that the effect of any pedestrian $A$ on pedestrian $B$ is absolutely independent of any other pedestrian in a scenario.

Superposition is easiest noticed for the Social Force Model due to the similarities with Newtonian Mechanics, however, if one thinks of superposition of {\em effects} not only forces, then many models with a different modeling approach are just as well superposing effects of pedestrians. 

\begin{figure}[htbp]
  \center
	\includegraphics[width=0.612\textwidth]{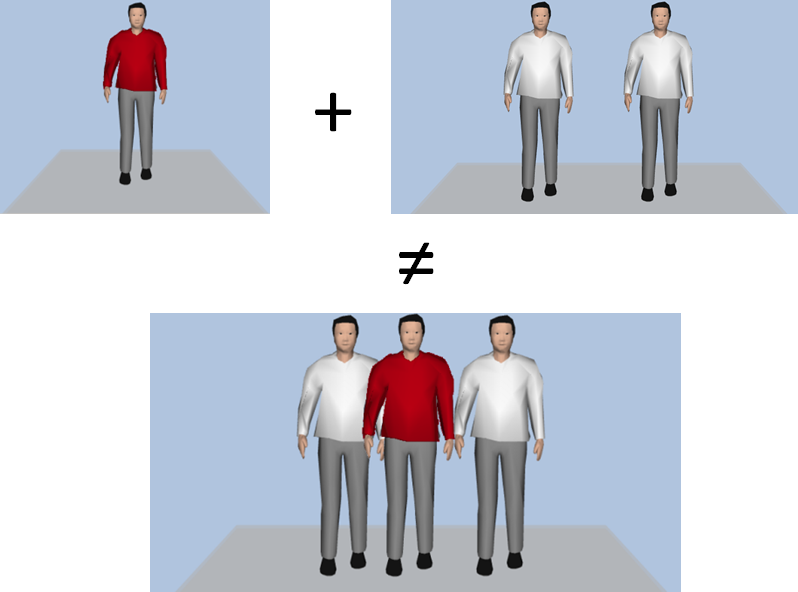} 
	\caption{An inequality with regard to the {\em effects} of the displayed situations. The idea is that the observer here basically wants to move straight on, but is faced with pedestrians walking the opposite direction. The two white pedestrians alone would eventually ``channel'' the observer to pass between them. In a force-based model their deviation triggering effects -- at least for this very moment -- would largely cancel. The red pedestrian alone would make the observer do a small detour to the left or to the right. However, together the three trigger the observer to do a larger detour to the left or to the right. The effect of all three -- especially the red one -- on the observer is determined by their membership in the total functional system.}
	\label{fig:holisitc}
\end{figure}

In physics superposition is so common that one might even wonder, how else a model should be constructed. One approach to the answer is to formulate the extreme opposite: ``the effect of any pedestrian $A_i$ onto a specific pedestrian $B$ is depends on all other pedestrians $A_j$ and their states''. This phrase is a concretion for (the simulation of) pedestrians of Sherif's general statement on social systems: ``the properties of any part are determined by its membership in the total functional system'' \cite{Sherif1936} and it is a {\em holistic} position. Figure \ref{fig:holisitc} shows a simple situation, where a superposition of effects does not yield a fully realistic effect and where therefore applying a holisitc approach improves the results.

The module of travel time estimation by construction is a holistic approach. One single pedestrian either can have a negligible effect in one setting and a very large one in a different setting, depending on the distribution and dynamic properties of all other pedestrians (and obstacles) in the system, even if his position and dynamic properties are the same in both settings. Therefore the holistic travel time estimation module is an ideal supplement to the superposing force approach.

This line of argumentation shall not mislead to the conclusion that a holistic (concrete: travel time estimating) approach alone is sufficient or superior to a superposing (concrete: force-based) approach. There are situations which are better addressed with forces. Imagine for example in Figure \ref{fig:holisitc} that there are walls to the left and the right of the group of three pedestrians. Then the observer would be {\em forced} to slow down, even be {\em forced} to stop. This is better modeled with {\em forces}. The travel time estimation module then is not of much help, as it is there to compute a good walking direction. Confined between two walls there would be no more really good direction, but just the task to slow down and resolve the situation without collisions. 

A different example which fits better for force-based modeling is when the density is high enough that each pedestrian is {\em forced} to go with the flow.

On the contrary scenarios which are only sparsely populated with individual pedestrians can be addressed with both approaches alike.

It is possible to some extend to use only one of the two approaches to model effects which are better modeled with the other approach (only forces and a simple determination of the desired walking direction as has been done usually or only a holistic approach and only a basic treatment of forces and accelerations \cite{Rascle2012}), but it makes life easier if both are applied, each for its own ``natural'' purpose.

With the emphasis in the previous lines it has already been insinuated: in a nutshell it appears that it is more natural and therefore easier to use {\em forces} to model behavior which is experienced to be {\em forced}\footnote{Note that not in every language the same word is used for physical force and an (en)forced action. In German it is for example ``Kraft'' and ``Zwang''.}. A main reason for this experience often is the small time available to react on a stimulus which does not allow for decent planning. On the contrary the holistic approach with its potentially unlimited range for strong interactions is better suited to model conscious planning, intentions or desires.

\section{The Effect of Integrating Travel Time into Models of Pedestrian Dynamics}
\subsection{Example 1}
\begin{figure}[htbp]
  \center
	\includegraphics[width=0.618\textwidth]{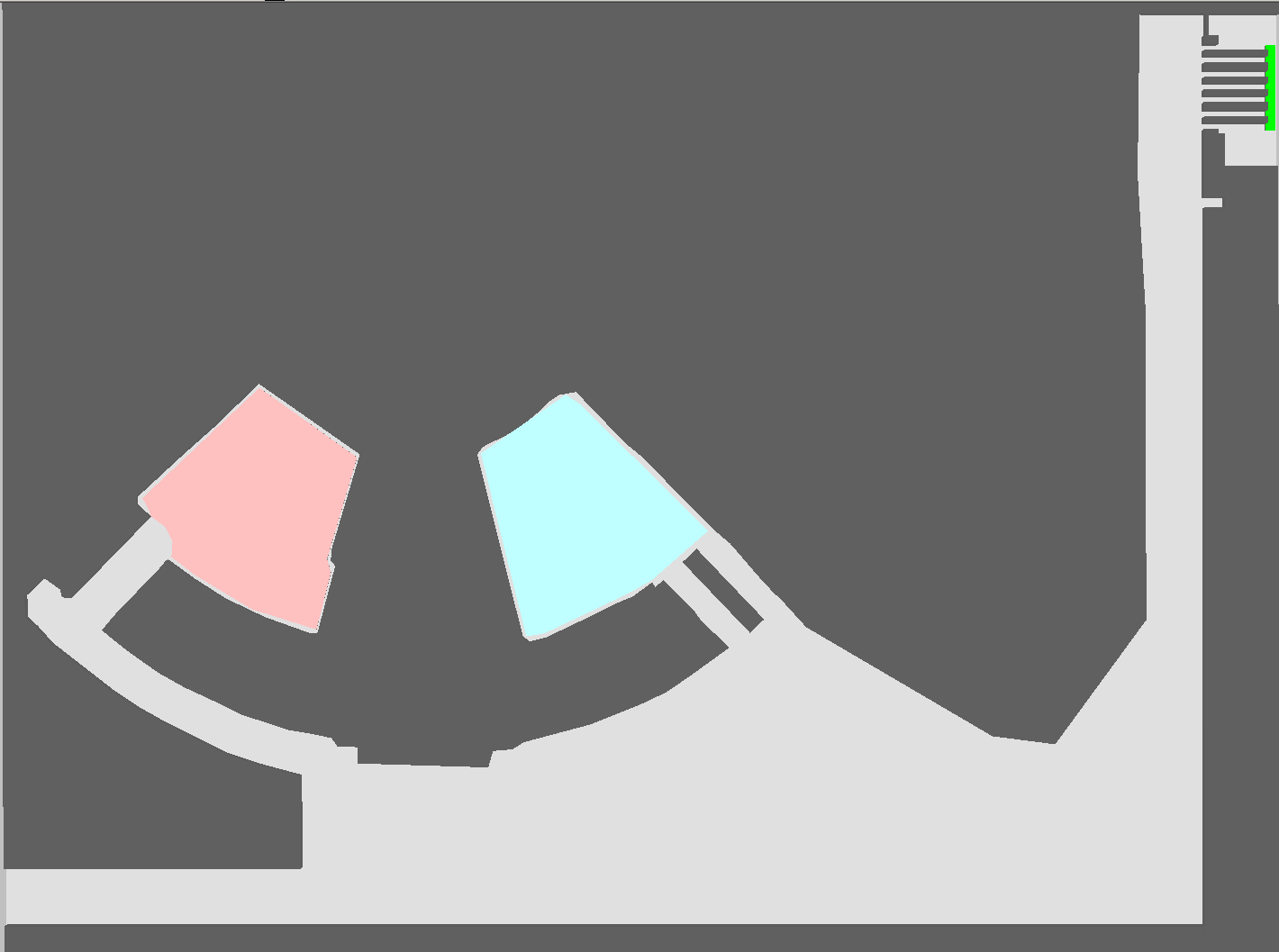}
	\caption{Scenario. Pedestrians can move on light grey areas, while dark grey areas denote obstacles. Pedestrians are set on the light red and light blue areas into the scenario and have to move to the green area at the upper right.}
	\label{fig:geometry}
\end{figure}

The scenario as shown in Figure \ref{fig:geometry} is simulated both with the F.A.S.T. model and the Social Force Model of PTV Viswalk {\em without} applying their modules of estimated remaining travel time. The geometry mimics a simplified part of a stadium and its exterior. In the beginning two groups with each 5,000 pedestrians start their movement in the interior of the stadium.

The first observation is that it is possible to find parameters for both models with which the simulation results become very similar first for the total evacuation times, second for the average individual egress time and third for the spatial en gros distribution of pedestrians all throughout the simulation. If one has a detailed look at the trajectories, these look differently for both models. However, the idea here is demonstrating equivalence for someone interested in an overview, i.e. envisioned as ``looking from high above on the whole scenario''. 

These are the parameters that were used with the F.A.S.T. model: $k_S=1.0$, $k_D=0.0$, $k_I=k_W=0.6$, $\mu=0.0$; and these with PTV Viswalk: $\tau=0.41$, $A_{soc,iso}=1.0$, $B_{soc,iso}=0.2$, $\lambda=0.0$, $A_{soc,mean}=0.4$, $B_{soc,mean}=2.8$, $VD=1.7$, $noise=1.2$. The latter parameters are identical or close to the default values except for $A_{soc,iso}$, whose value has been clearly reduced to achieve the relatively high densities which occur in the F.A.S.T model with a comparable regularity. 

In both cases the free (or desired or preferred) speeds have been (nearly) equally distributed between 1.2 and 2.0 m/s. As they are defined with different methods in both models, the distribution of free speeds is be not exactly identical, however very similar. Note that it has been found to be crucial for the equivalence of the results of both approaches to have a very similar distribution of free speeds.

Someone familiar with one of the models will immediately see that neither the chosen parameters nor the free speeds are extreme in any way.

With these settings the individual egress and the total evacuation times result as shown in Table \ref{tab:results1}.

\begin{table}[htbp]
	\center
	\begin{tabular}{|l|ccc|ccc|} \hline
		       &\multicolumn{3}{|c|}{Av. Individual Egress Time}&\multicolumn{3}{c|}{Total Evacuation Time}\\
		       &   Min., &   Average $\pm$ STD,  &   Max.   &   Min., &   Average $\pm$ STD,  &   Max. \\ \hline
	F.A.S.T. &  1317.1 &   1332.2  $\pm$ 4.9   & 1347.9   &  2529.0 &   2558.9  $\pm$ 9.3   &  2590.0\\
	Viswalk  &  1324.0 &   1337.4  $\pm$ 4.6   & 1350.6   &  2528.2 &   2555.9  $\pm$ 9.1   &  2584.3\\ \hline
	\end{tabular}
	\caption{Results statistics from 1,000 simulation runs with each of the models. ``Av. Individual Egress Time'' means the average of egress times of all pedestrians of one simulation run. Over 1,000 runs there are 1,000 averages of which the minimum, average, and maximum are displayed.}
	\label{tab:results1}
\end{table}

\begin{figure}[htbp]
  \center
	\includegraphics[height=0.17\textheight]{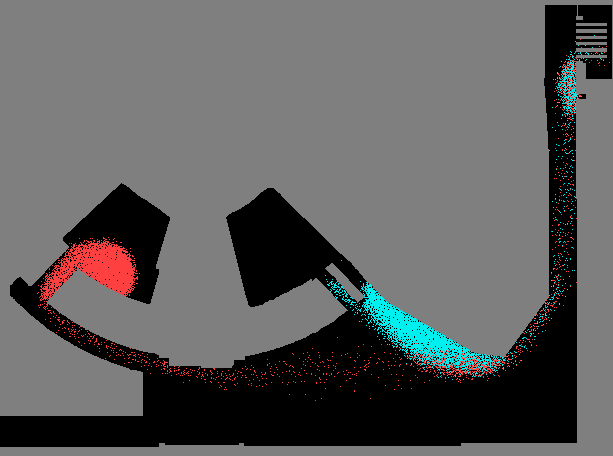} \hspace{12pt}
	\includegraphics[height=0.17\textheight]{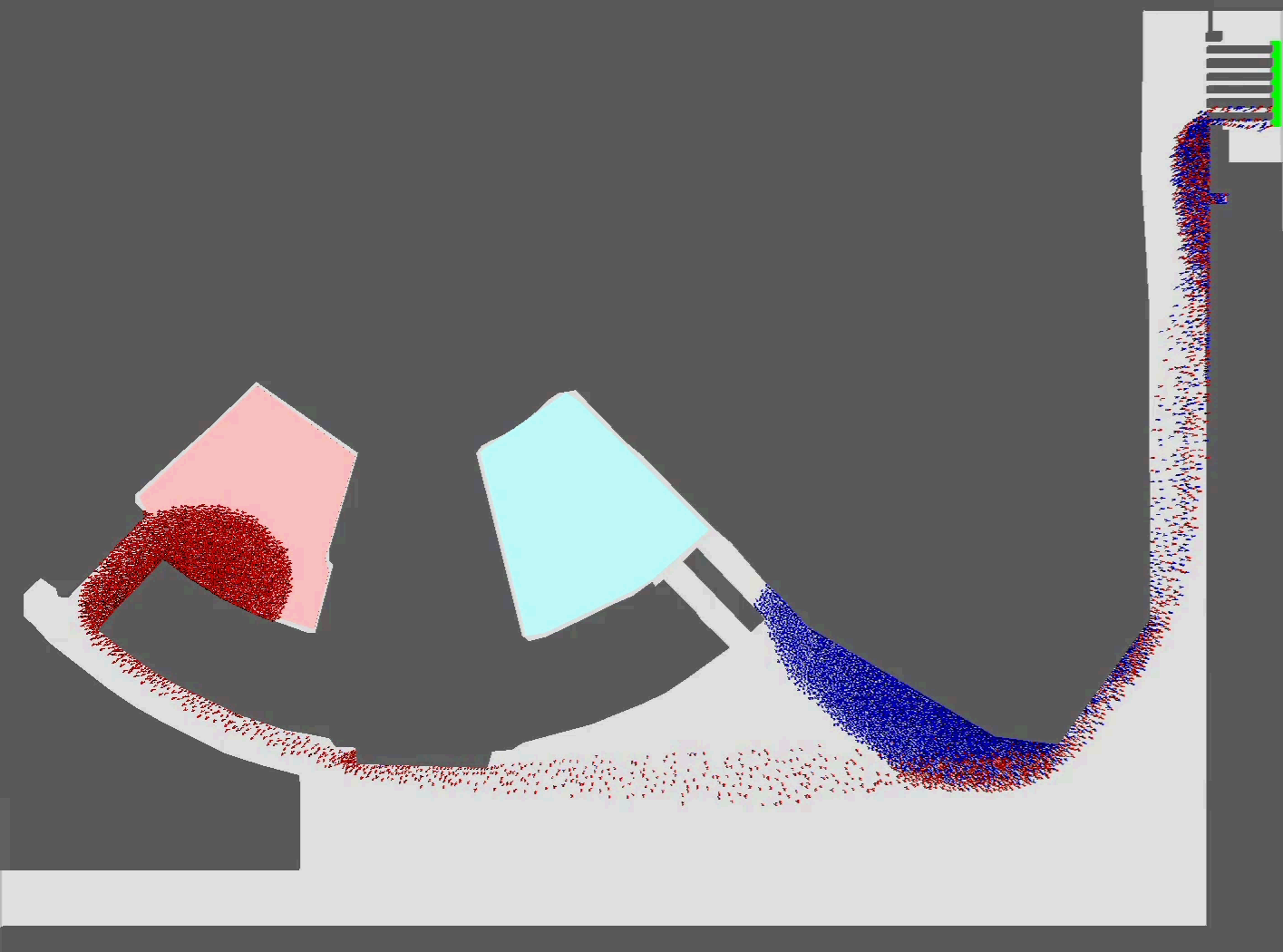} \\ \vspace{1pt}
	\includegraphics[height=0.17\textheight]{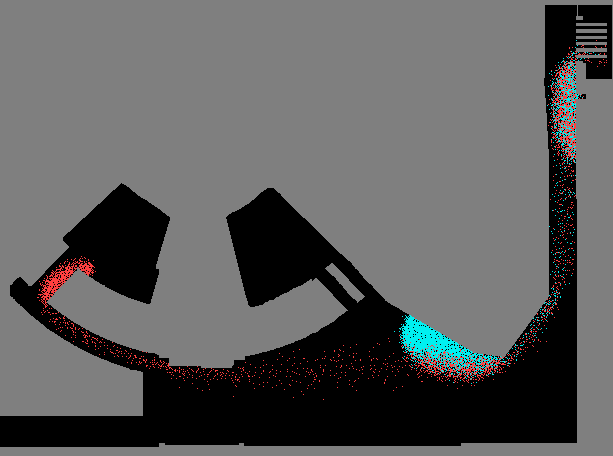} \hspace{12pt}
	\includegraphics[height=0.17\textheight]{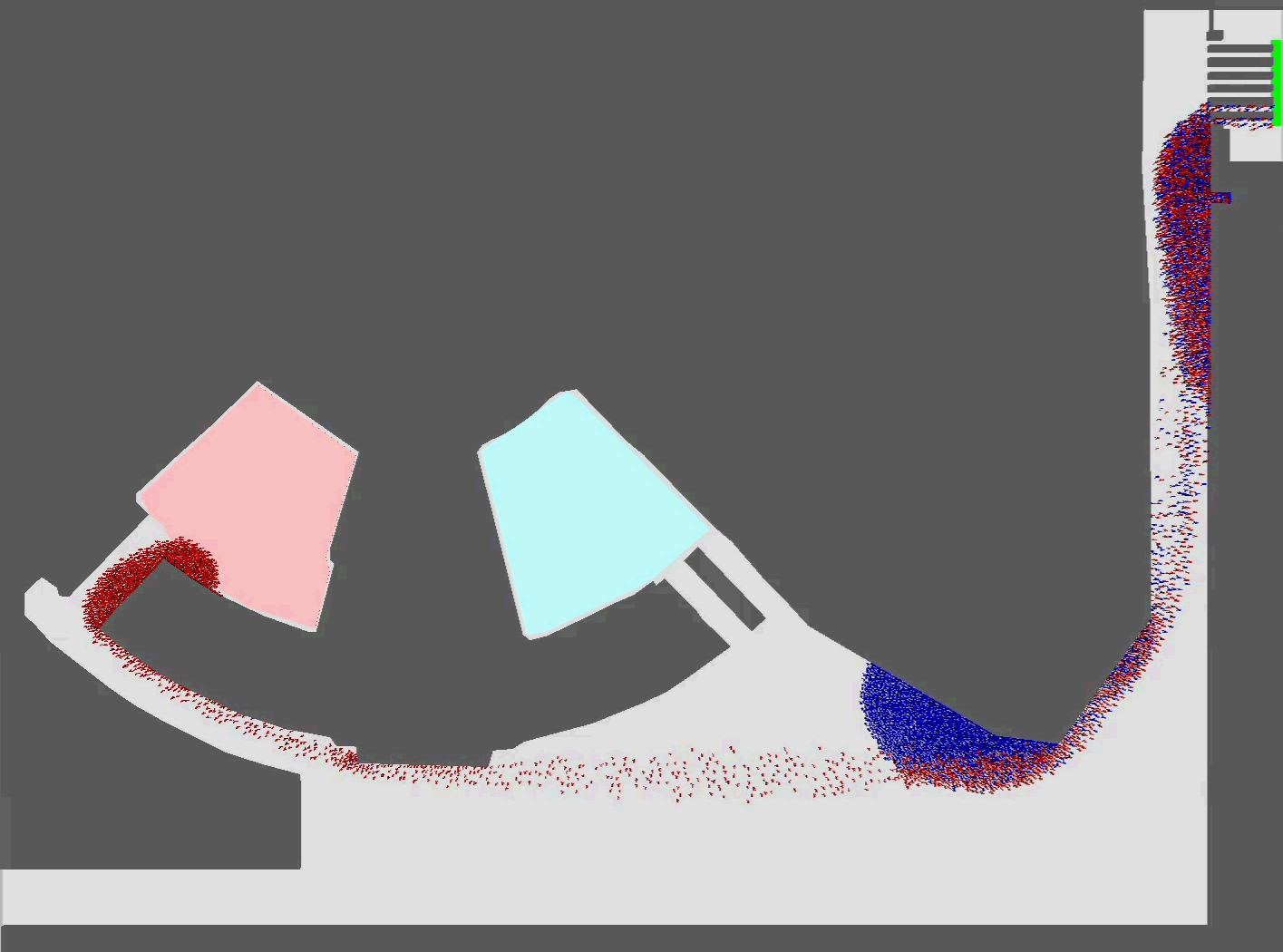} \\ \vspace{1pt}
	\includegraphics[height=0.17\textheight]{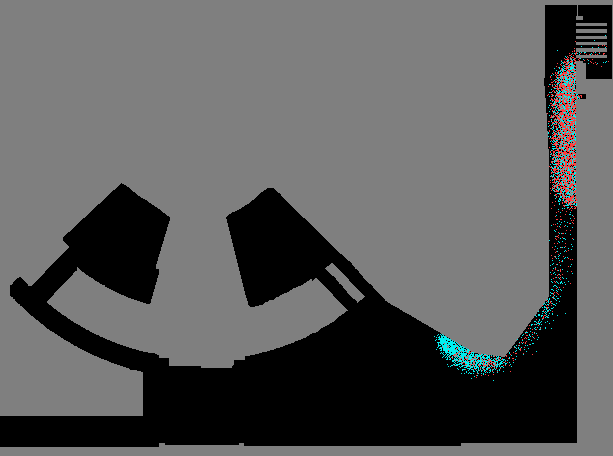} \hspace{12pt}
	\includegraphics[height=0.17\textheight]{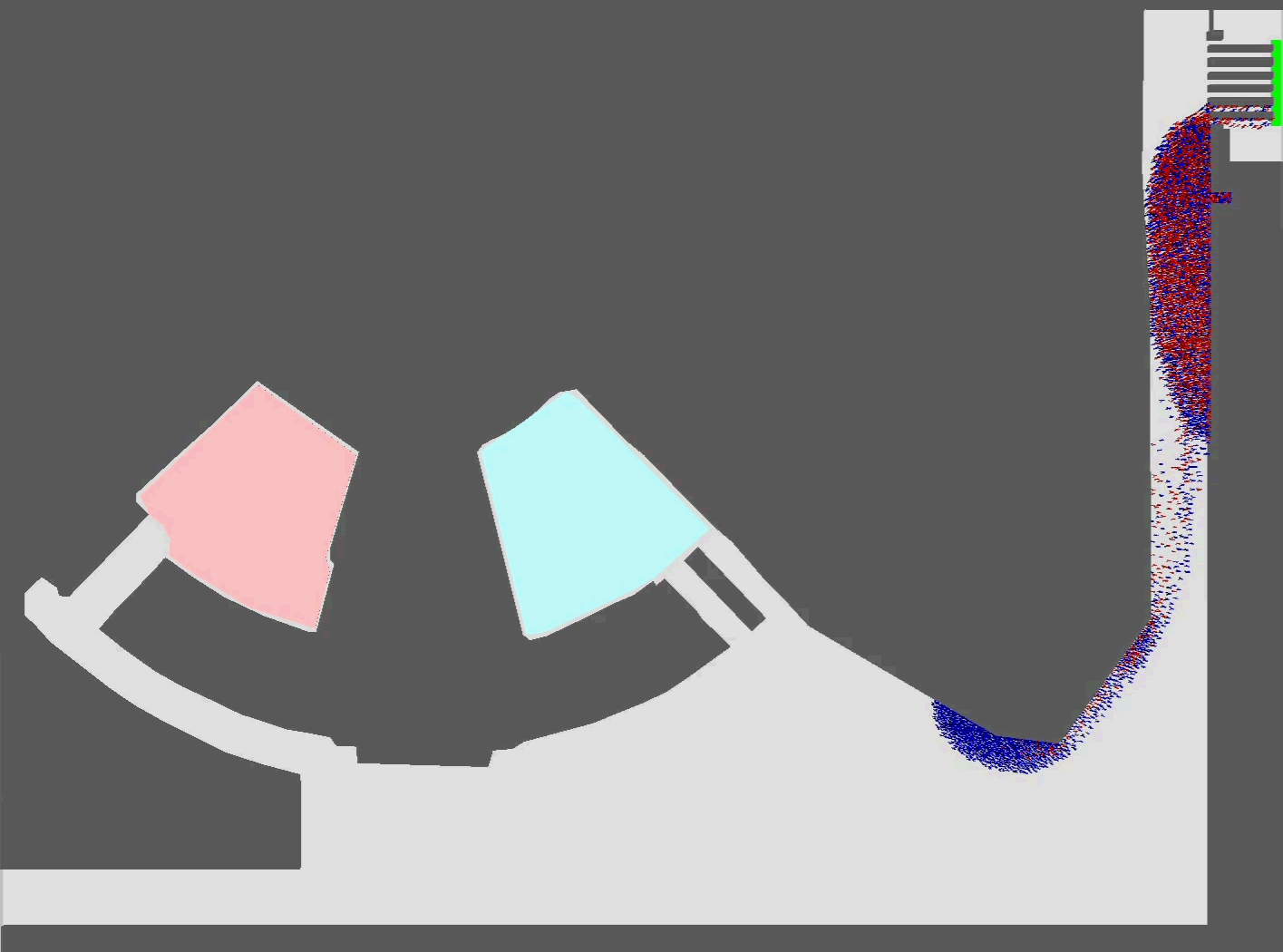} \\ \vspace{1pt}
	\includegraphics[height=0.17\textheight]{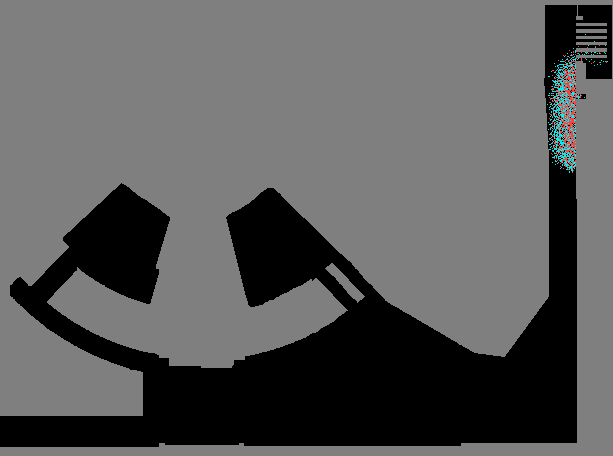} \hspace{12pt}
	\includegraphics[height=0.17\textheight]{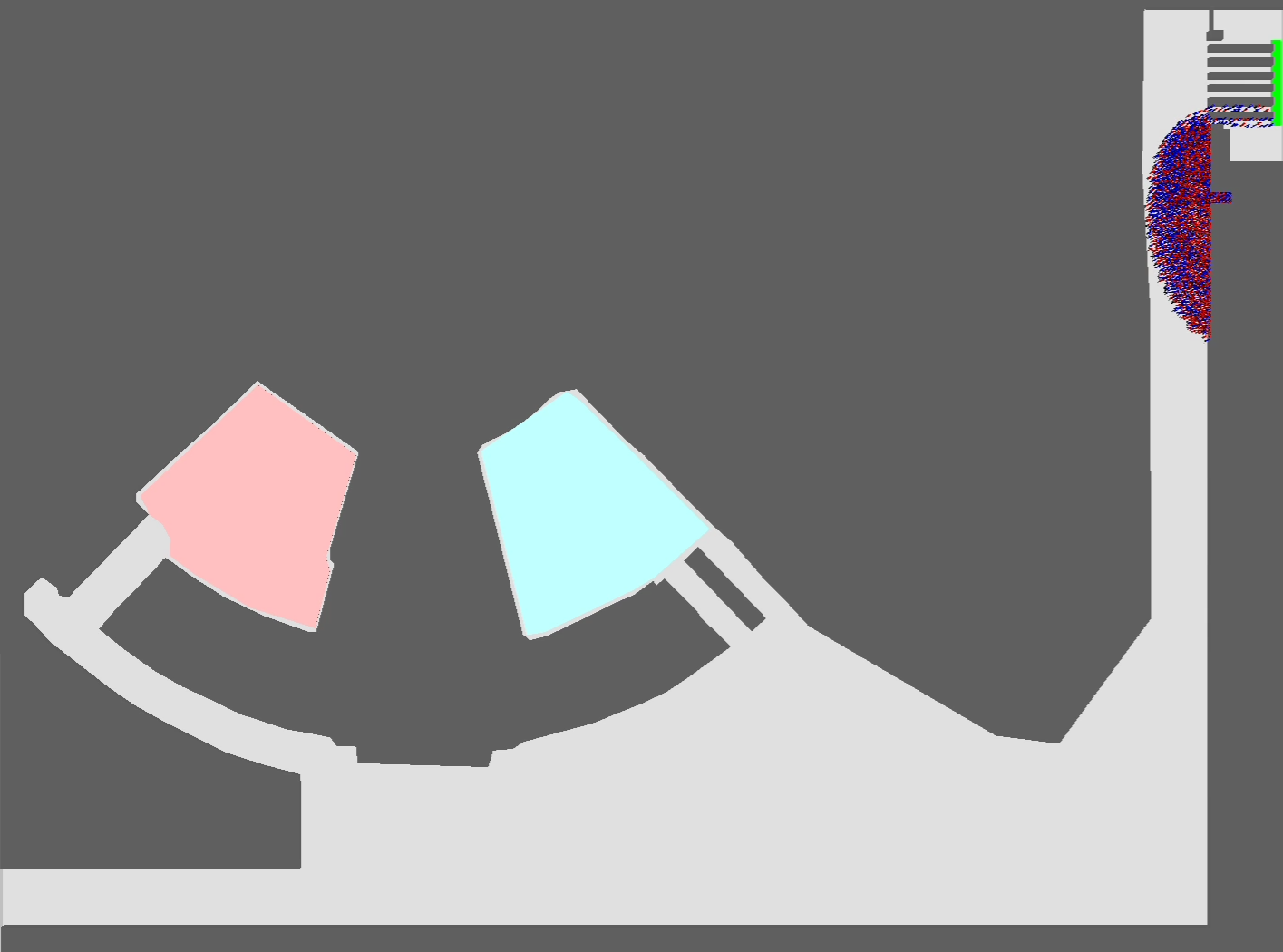} \\ \vspace{1pt}
	\includegraphics[height=0.17\textheight]{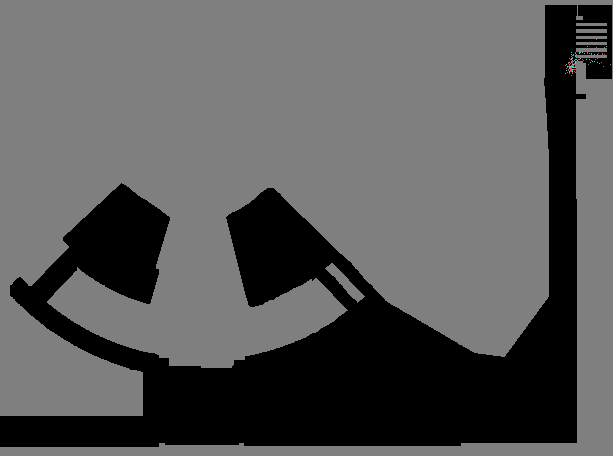} \hspace{12pt}
	\includegraphics[height=0.17\textheight]{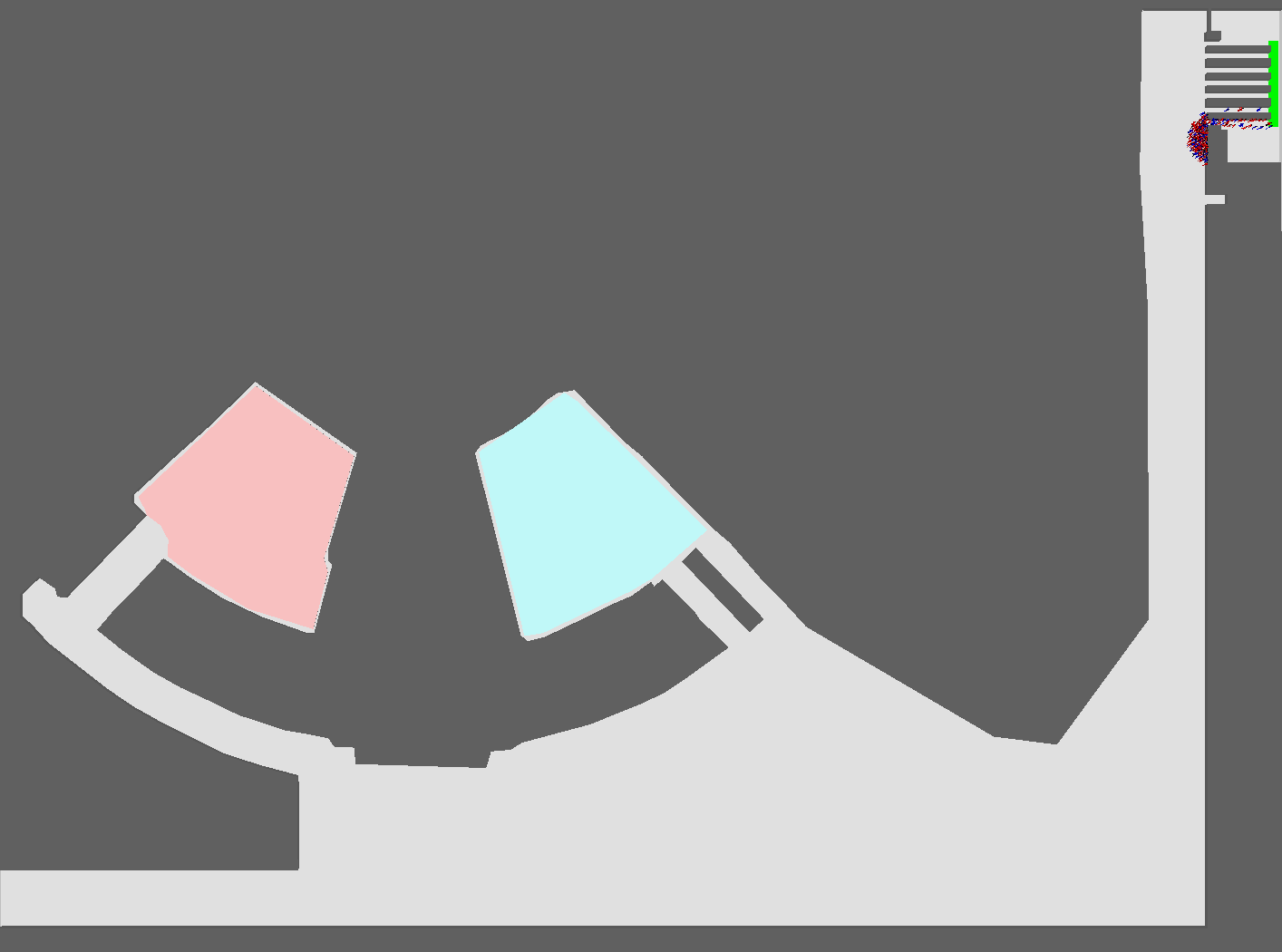} \\
	\caption{Snapshots from the simulations without dynamic potential: left: F.A.S.T., right: Viswalk, at times 500 sec., 1000 sec., 1500 sec., 2000 sec., and 2500 sec..}
	\label{fig:screenshots}
\end{figure}

Selecting for both models the simulation run which has the least sum of mean square deviations from the averages for total evacuation and individual egress times Figure \ref{fig:screenshots} in this way compares the average simulation runs of both models and shows that the results are very similar all through time and space.

While these snapshots show how similar the results are, a second look reveals that the pedestrians move unrealistically close around the corners. As a consequence they ignore all but two or three of the exit gates shortly before the destination area. One may therefore expect that the times as shown in Table \ref{tab:results1} are much larger than would be realistic. Someone familiar with one or both of the models (better: model classes) will know that this cannot be fixed by just modifying the parameters. Of course simulated egress and evacuation times can be reduced to some degree with different parameters, but only at the cost of having the models yield unrealistic results somewhere else in this geometry and even more so in different geometries. A reader not familiar with these models here needs to trust the author that this is in fact the case as demonstrating this would by far exceed the available space of this contribution and a strict formal prove is simply not possible.

Using the module of remaining travel time estimation on contrast greatly reduces egress and evacuation times and strongly influences the walking behavior in both models. For these simulations the same parameters as above have been used and in addition for the travel time estimation module used with the F.A.S.T. model it has been set $k_{Sdyn}=1.0$ and $s_{add}=10$ (see \cite{Kretz2009,Kretz2009c}) and for the one of PTV Viswalk $g=1.5$, $h=0.7$ and the impact strength has been set to 100\% (see \cite{Kretz2011e}). For this second stage of simulations it has not been attempted actively to find parameters with which the results of the two approaches match, i.e. these values are something like the defaults of the methods judged by previous experience.

\begin{table}[htbp]
	\center
	\begin{tabular}{|l|ccc|ccc|} \hline
		       &\multicolumn{3}{|c|}{Av. Individual Egress Time}&\multicolumn{3}{c|}{Total Evacuation Time}\\
		       & Min., & Average $\pm$ STD, & Max.  & Min.,  & Average $\pm$ STD, & Max.   \\ \hline
	F.A.S.T. & 784.6 &  787.6  $\pm$ 1.8  & 789.9 & 1345.0 & 1352.7  $\pm$ 4.9  & 1363.0 \\
	Viswalk  & 666.5 &  672.0  $\pm$ 2.4  & 676.2 & 1227.4 & 1234.9  $\pm$ 4.6  & 1244.2 \\ \hline
	\end{tabular}
	\caption{Results statistics from 15 simulation runs with each of the models when the travel time estimation module is used.}
	\label{tab:results2}
\end{table}

\begin{figure}[htbp]
  \center
	\includegraphics[width=0.45\textwidth]{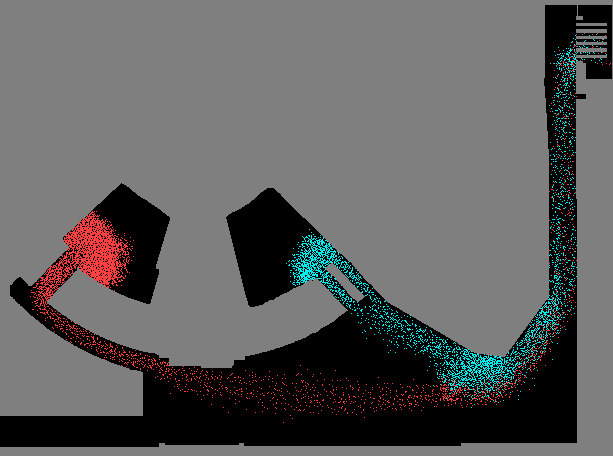} \hspace{6pt}
	\includegraphics[width=0.45\textwidth]{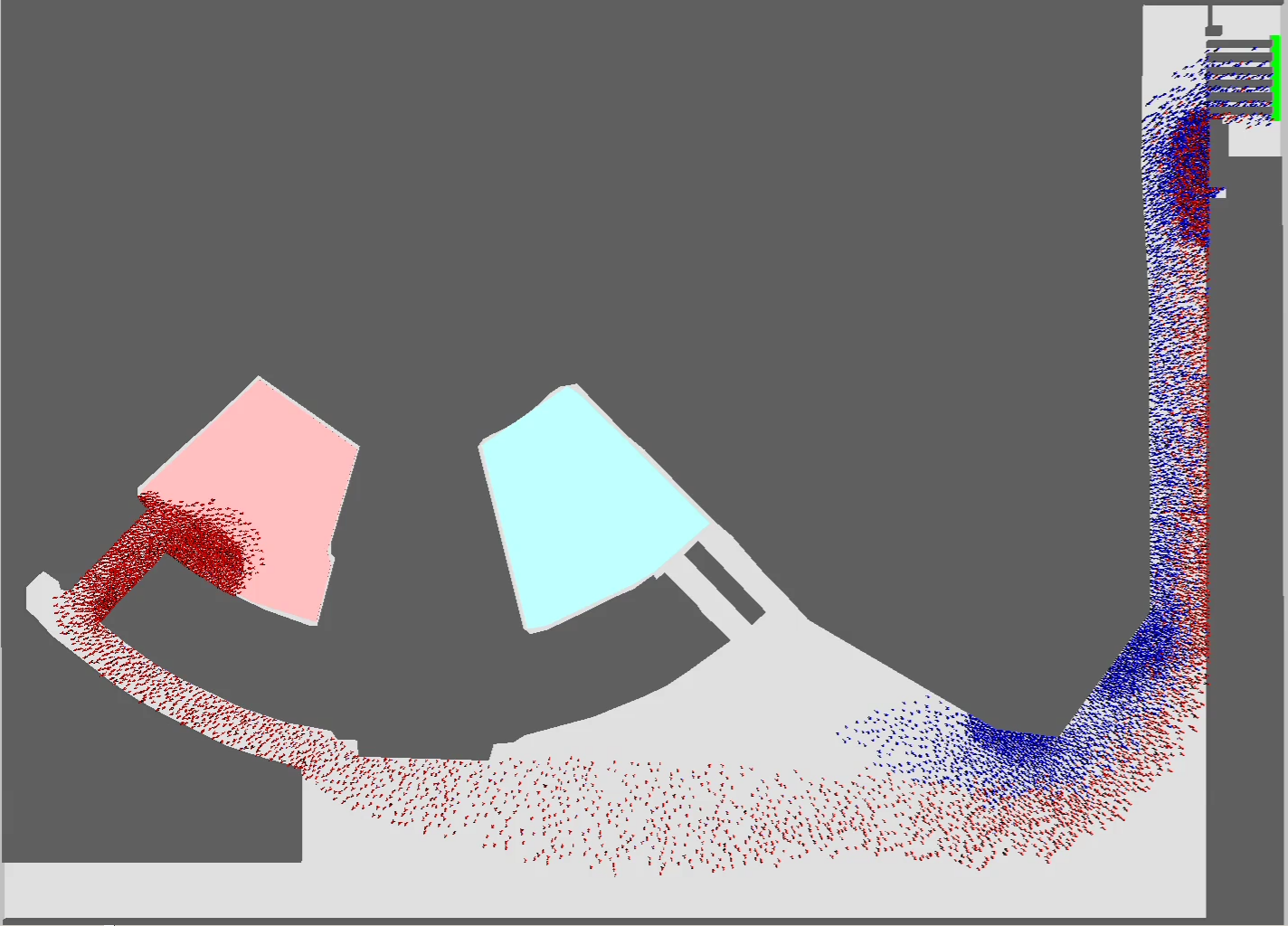} 
	\caption{Snapshots from the simulations with dynamic potential: left: F.A.S.T., right: Viswalk, each at time 350 seconds.}
	\label{fig:screenshots2}
\end{figure}

Figure \ref{fig:screenshots2} shows snapshots from both simulations.  By comparing with Figure \ref{fig:screenshots} it can be seen that the travel time estimation module modifies the simulation dynamics of both models into the same direction: pedestrians utilize the available space at corners better, they utilize the full width of the long corridor to the right better and they utilize the exit gates better compared to the simulations without travel time estimation method as shown in Figure \ref{fig:screenshots} and the red pedestrians -- respectively a center line through the stream of red pedestrians -- in the open area do not head for the corner of the obstacle but (about) for the edge of the group of blue pedestrians. Last but not least: for both models the evacuation and egress times are about halved as a comparison of Tables \ref{tab:results1} and \ref{tab:results2} shows.

\subsection{Example 2}
\begin{figure}[htbp]
  \center
	\includegraphics[width=0.612\textwidth]{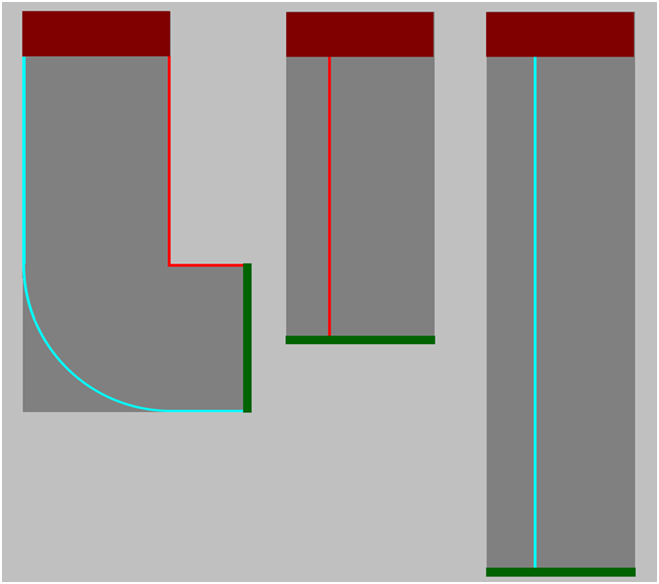}
	\caption{The corner and the two straight geometries for comparison. 500 pedestrians have to walk from the dark red to the dark green area. The red line is 38 m (28 + 10) long, the cyan one 69.4 m (28 + 10 $\pi$ + 10) long. The width of the areas is 20 m. The dark red insertion area has a length of 6 m.}
	\label{fig:example2}
\end{figure}

The first example was composed of a number of basic elements: corridors, corners, gates. To clarify the effect of the travel time modules this is now to be amended by a most simple example: a group of pedestrians walks around a corner and this is compared with examples where the same number of pedestrians walks comparable distances straight on. Details of the geometry are shown in Figure \ref{fig:example2}

The corner situation is simulated once without and once with activated travel time estimation module. For the straight examples the travel time estimation is not used it would not have much effect there anyway. So why simulate it at all? The straight geometries are only there to set a reference frame which makes the results less dependent on specific parameter settings. The value of the simulations in the straight geometries will become apparent with the results.

The results for the F.A.S.T. model are given in Table \ref{tab:results3} and for PTV Viswalk in Table \ref{tab:results4}. In this second example it was not intended to achieve agreement between both models. Instead PTV Viswalk was run with default parameters and for the desired speed a population composition was chosen as in the RiMEA \cite{Rimea2009} test cases (50\% male, 50\% female; 4\% mobility impaired and beyond that the age groups ($<$30, 30-50, and $>$50) with equal ratios) with their desired walking speeds according to \cite{Weidmann1993}. In this way much smaller desired speeds were involved than for the F.A.S.T. model where preferred speeds between 1.2 and 2.0 m/s were used. The other parameters were set as in example 1.

\begin{table}[htbp]
	\center
	\begin{tabular}{|l|ccc|ccc|} \hline
		             &\multicolumn{3}{|c|}{Individual Egress Time}&\multicolumn{3}{c|}{Total Evacuation Time}\\
		             & Min., & Average $\pm$ STD, & Max.  & Min., & Average $\pm$ STD, & Max. \\ \hline
	Corner w/o     & 79.3  & 84.0    $\pm$ 1.2  & 88.6  & 142.0 & 154.2  $\pm$ 3.4  & 171.0 \\
	Corner w       & 66.2  & 69.4    $\pm$ 0.8  & 72.6  & 109.0 & 114.8  $\pm$ 2.3  & 127.0 \\
	Straight short & 41.8  & 43.4    $\pm$ 0.5  & 45.0  &  65.0 &  70.1  $\pm$ 1.9  &  80.0 \\
	Straight long  & 67.2  & 69.8    $\pm$ 0.8  & 72.5  & 102.0 & 107.4  $\pm$ 2.2  & 118.0 \\ \hline
	\end{tabular}
	\caption{Results statistics from 3000 simulation runs with the F.A.S.T. model. The first line gives the times when no travel time estimation module is used, the second when it is.}
	\label{tab:results3}
\end{table}

\begin{table}[htbp]
	\center
	\begin{tabular}{|l|ccc|ccc|} \hline
		             &\multicolumn{3}{|c|}{Individual Egress Time}&\multicolumn{3}{c|}{Total Evacuation Time}\\
		             & Min., & Average $\pm$ STD, & Max.  & Min., & Average$\pm$ STD, & Max.  \\ \hline
	Corner w/o     & 104.2 & 110.5   $\pm$ 1.9  & 117.0 & 229.4 & 248.1  $\pm$ 6.6  & 274.4 \\
	Corner w       &  78.2 & 82.5    $\pm$ 1.4  & 88.3  & 157.9 & 174.2  $\pm$ 5.0  & 191.9 \\
	Straight short &  52.0 & 55.1    $\pm$ 0.9  & 58.3  &  96.6 & 111.1  $\pm$ 3.4  & 123.2 \\
	Straight long  &  81.9 & 86.6    $\pm$ 1.4  & 90.9  & 150.4 & 176.2  $\pm$ 5.4  & 188.8 \\ \hline
	\end{tabular}
	\caption{Results statistics from 3000 simulation runs with PTV Viswalk. The first line gives the times when no travel time estimation module is used, the second when it is.}
	\label{tab:results4}
\end{table}

Although two different models, two different methods for travel time estimation, and different distributions of the desired speed were used, and although for this reason the total evacuation time and the individual average egress time differ in their absolute value for F.A.S.T. and PTV Viswalk, the results shown in Tables \ref{tab:results3} and \ref{tab:results4} are similar with regard to a number of properties (numbers with regard to average values): 
\begin{list}{--}{}
\item The individual average egress time in the corner geometry is 120\% (F.A.S.T) respectively 128\% (PTV Viswalk) compared to the case of the straight long corridor if no travel time estimation is made. However, it drops to 99\% respectively 95\%, if the travel time estimation module is active. Simpler spoken: they are clearly longer without and similar with travel time estimation. 
\item For the total evacuation time about the same relation holds: there the time drops from 144\% (F.A.S.T) respectively 141\% (PTV Viswalk) to 107\% respectively 99\%, if the travel time estimation module is active. Again the summary is that the time is clearly longer in the corner geometry compared to the straight long corridor if the travel time estimation module is not used, but it is about the same with both models if it is used.
\item With the corner geometry the average individual egress time is 83\% (F.A.S.T.) respectively 75\% (PTV Viswalk) when the travel time estimation module is used compared to if it is not used.
\item With the corner geometry the average total evacuation time is 74\% respectively 70\% when the travel time estimation module is used compared to if it is not used.
\item If the travel time estimation module is active the individual average egress time in the corner geometry is 160\% (F.A.S.T.) respectively 150\% (PTV Viswalk) of the one in the straight short corridor.
\item If the travel time estimation module is active the total egress time in the corner geometry is 164\% (F.A.S.T.) respectively 157\% (PTV Viswalk) of the one in the straight short corridor.
\end{list}

As the properties for F.A.S.T. and PTV Viswalk in all cases are separate by a maximum of ten percent points, while the effect of the travel time estimation module is much larger, these relations show that the details of the pedestrian model, its parameters (especially walking speeds) and the details of the travel time estimation module method and its parameters are of minor relevance. The effect is clear and distinguishes the case when it is used from the case when it is not used. This implies that with empirical data it should be clearly possible to decide if real people move in such a situation rather based on a shortest or a quickest path paradigm, i.e. if it enhances the degree of realism when the travel time estimation module is used. In addition it should be possible to compute those relations above where times at the corner geometry are related with times in one of the straight corridors with any pedestrian simulation model available. If then for example total evacuation time at the corner is about 150\% of the one in the straight long corridor it can be concluded that pedestrians in the model navigate according to a shortest path paradigm, while if the times are about equal or the time in the corner geometry is even shorter than the one in the long straight corridor it can be conclued that they follow a quickest path paradigm.

\begin{figure}[htbp]
  \center
	\includegraphics[width=0.2\textwidth]{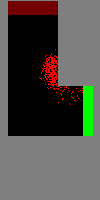} \hspace{6pt}
	\includegraphics[width=0.2\textwidth]{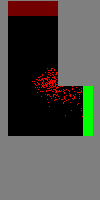} \hspace{6pt}
	\includegraphics[width=0.2\textwidth]{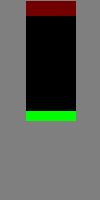} \hspace{6pt}
	\includegraphics[width=0.2\textwidth]{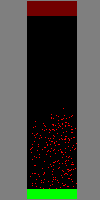}
	\caption{Snapshots from the simulations with the F.A.S.T. model at 70 seconds.}
	\label{fig:screenshots3}
\end{figure}

\begin{figure}[htbp]
  \center
	\includegraphics[width=0.8\textwidth]{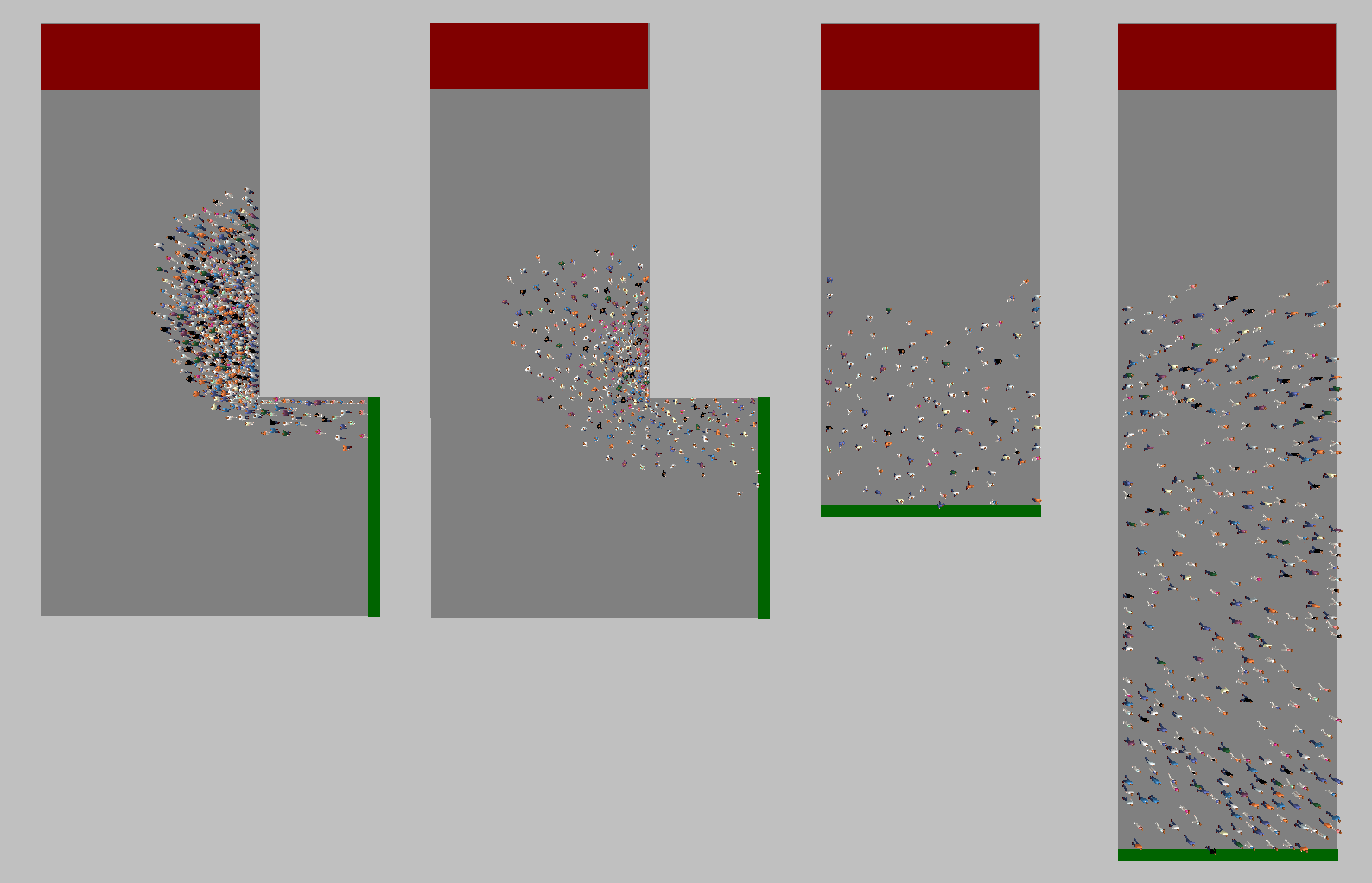} 
	\caption{Snapshots from the simulations with Viswalk at 70 seconds.}
	\label{fig:screenshots4}
\end{figure}

For completeness Figures \ref{fig:screenshots3} and \ref{fig:screenshots4} show snapshots from the simulations.

\section{Summary}
In this contribution it was argued that it is of greater relevance for the degree of simulation results of a pedestrian simulation model that pedestrians walk into the direction of estimated earliest arrival instead of the direction of the shortest path than the details of a) the model of pedestrian dynamics and b) the details of the travel time estimation module. For this it was first shown that in a non-trivial example two different approaches to modeling pedestrian dynamics can yield very similar results on a moderate level of aggregation if one sets the model parameters appropriately. It was then shown that for both approaches the degree of realism is increased, if one changes from a shortest path direction to a quickest path direction paradigm.

The examination of a simple example scenario confirmed this. With two different models and different desired walking speed distributions and different travel time estimation modules it was found that the effect of using a travel time estimation module (or using it not) had about the same impact on certain relative properties. This implies a comfortable situation for comparison with empirical data.

A first obvious conclusion is that it would be very interesting to have empirical data for situations as the ones in example 2: where between the results for the two straight corridors do the results of the corner geometry lie for real people?

The second conclusion is that because it is a) possible to have some cellular automata-based model yield such similar results and b) not possible for both modeling approaches to enhance the degree of realism in certain situations, variants of the Social Force Model which modify exclusively or mainly the forces and do not include a travel time estimation module \cite{Lakoba2005,Yu2005,Chraibi2010} most probably cannot help to improve the realism of results to the same degree as the travel time estimation module can.

Finally one can summarize that in the formulation of a pedestrian simulation model we can let guide ourselves by a quote which is ascribed to Blaise Pascal: {\em ``Desire {\bf and}\footnote{Emphasize added by the author of the facing work.} Force are responsible for all our actions.''} A general way to interpret it in this context is that one equally has to take care for desired (planned or conscious) and for enforced (reactive, unplanned) determinants of motion. A more concrete way to interpret it for a force-based approach to pedestrian dynamics is that one has to put equally much effort in modeling the desired velocity as in modeling the forces.

%
\nocite{_Ency2009book,_ACRI2010}
\bibliographystyle{utphys2011}
\bibliography{PED-DDPF}
%
\end{document}